\documentclass[a4paper,UKenglish]{lipics-v2018}

\usepackage{microtype}

\title{Information carefull worstcase \Decrement\ heaps with simple non\Meld\ variant}

\author{Vladan Majerech}{Department of Theoretical Computer Science and Mathematical Logic (KTIML), Charles University, Malostransk\'e n\'am\v est\'\i\ 25, Prague 118 00, Czech Republic}{maj@ktiml.mff.cuni.cz}{ https://orcid.org/0000-0003-3006-2002}{}

\authorrunning{V. Majerech}

\Copyright{Vladan Majerech}

\subjclass{Information systems$\rightarrow$Information storage systems$\rightarrow$Record storage systems$\rightarrow$Record storage alternatives$\rightarrow$Heap (data structure)}


\category{}

\relatedversion{}

\supplement{}

\funding{}

\acknowledgements{}

\EventEditors{}
\EventNoEds{1}
\EventLongTitle{}
\EventShortTitle{}
\EventAcronym{}
\EventYear{2019}
\EventDate{}
\EventLocation{}
\EventLogo{}
\SeriesVolume{}
\ArticleNo{1}
\nolinenumbers 
\hideLIPIcs  

\begin{document}
\newcommand{\prevfootnotemark}{\csname @footnotemark\endcsname}
\newcommand{\MakeHeap}{{\bf MakeHeap\/}}
\newcommand{\FindMin}{{\bf FindMin\/}}
\newcommand{\DeleteMin}{{\bf DeleteMin\/}}
\newcommand{\Delete}{{\bf Delete\/}}
\newcommand{\Meld}{{\bf Meld\/}}
\newcommand{\Cut}{{\bf Cut\/}}
\newcommand{\Insert}{{\bf Insert\/}}
\newcommand{\Decrement}{{\bf DecreaseKey\/}}
\newcommand{\Right}{\hbox{\sl right\/}}
\newcommand{\Left}{\hbox{\sl left\/}}
\newcommand{\nill}{\hbox{\sl null\/}}
\newcommand{\Parent}{{\sl parent\/}}
\newcommand{\numfootnote}{\footnote}

\maketitle

\begin{abstract}
We analyze priority queues including \Decrement\ method in its interface. The paper is inspired by Strict Fibonacci Heaps \cite{StrictHeaps}, where G. S. Brodal, G. Lagogiannis, and R. E. Tarjan implemented the heap with \Decrement\ and \Meld\ interface in assymptotically optimal worst case times (based on key comparisons). 
At the end of the paper there are mentioned possible variants of other structural properties an violations than they have used in the analysis.

In the main variant a lot of information is wasted during violation reduction steps. 
Our goal is to concentrate on other variants and to invent natural strategy not losing that much in the information value.
In other words we try to choose among them one which corresponds to superexpensive comparision principle as much as possible.
The principle was described in \cite{PadovanHeaps} of myself, but after publication I have found these ideas in \cite{FibonacciHeapsRevisited} of H. Kaplan, R. E. Tarjan, and U. Zwick. 
\end{abstract}

\section{Introduction}
Actually the worst case bounds were previously achieved by G. S. Brodal \cite{BrodalHeaps} unfortunately with rather wide heap and complicated maintainance reflected in big multiplication constants making them impractical. I will call heaps from the paper \cite{StrictHeaps} BLT heaps as their connection to Fibonacci is only negligable.
The BLT heaps are narrow and they are trying to avoid unnecessary organization as much as possible. 
Our main goal is to implement \Decrement\ heaps interface. 

In Fibonacci Heaps \cite{FibonacciHeaps} the \Meld\ method is internally used so it's for free to make it in the heap interface.
On the contrary, \Meld\ is not natural for the BLT heaps so implementing it requires a bit of structural overhead. Authors of \cite{StrictHeaps} shown, it could be done without changing assymptotical behaviour of the data structure. We will show, our strategy is able to do it as well if really necessary.
I don't know an algorithm where \Meld\ operation is required so I expect for most applications the simplified interface and simplified implementation would be prefered choice.

In the BLT heaps are two main tricks to allow worst case complexities in the heaps.
One is concerning decrement of the heap size $n$ during sequence of $\DeleteMin$s where the time bounds should reflect current $n$.
When node neighbourhood in heap need to have shape bounded by a function $b(n)$, the decrease of $n$ means we should recheck all the nodes regularly.
They have introduced a global list of heap nodes (just heap root is held separately) and after each decrement of $n$, 
2 first nodes are rechecked and moved to the list end. As the bound is logarithmic ($\le c_1+c_2\log_2 n$), we could check all nodes before the bound is decremented by $c_2$.
They have found nice describtion of this by using $b(2n-p)$ rather to $b(n)$ for bounding function of $p$-th node on the list. With this trick the bound remain constant for all nodes except those we are checking and it is at least $b(n+1)$.
The bound for nodes we are checking changes from $c_1+c_2\log_2(2n-1)$ resp. $c_1+c_2\log_2(2n-2)$ to $c_1+c_2\log_2(2(n-1)-(n-3))$ resp. $c_1+c_2\log_2(2(n-1)-(n-2))$ having difference either $c_2(\log_2(2n-1)-\log_2(n+1))<c_2$ or $c_2(\log_2(2n-2)-\log_2(n))<c_2$.
So planning degreee reduction by at least $c_2$ and doing it to extend as possible would maintain the degree constraints.
We would need this trick for \Meld\ including version as well, but no such complication would be required for the simplified version.

The other trick is not trying to be too pedantic on the heap structure. 
They allow some violations of idealistic heap shape, but they remember the violations classified by their type and rank of node it affected.
Violations of each type could be easily solved if we have at least two nodes of the same rank with this type of violation.
So whenever operation introduces new violation of given type, violation reduction for this type is called. Either set of nodes with the type of violation is small or there are nodes of same rank allowing violation size reduction in constant time. 
Therefore this system maintains the violation size naturally bounded.
Actually solving one type of violation may introduce violation of another type, but there is a linear function with small integer multiplicative constants combining violations sizes of all types, such that each violation reduction step decrements the function value whenever violation sizes are above their equilibrium values.
As the function is growing only by a constant per increase in a coordinate, only constant number of violation reduction steps is required (and could be easily planned) to maintain the violation sizes of all the violations not exceeding their equilibrium values.

The BLT paper needed to show the ranks are limited by some function $R(n)$ when each violation size is bounded by $R(n)+1$ and $R(n)\in O(\log n)$.
They got $R(n)$ be the bigger root of quadratic polynomial defined by $x=\log n+2+\sqrt{2(x+1)}$, it is easily bounded by $R(n)\le 6+2\log_2 n$, but even $R(n)\le 6+1.2\log_2 n$ would suffice. The same bound would apply to our heaps.
Using estimate $R(n)\le 6+2\log_2 n$ so $c_2=2$ gives plan to reduce degree of checked nodes by 2 for BLT heaps.

\section{Analysis of worst case heaps requirements}
Unfortunately, word active is used in citations in three different meanings, so we rather decided not use the word active in connection with our heaps at all. 
This is why I would use deffered/solid in meaning pasive/active from \cite{StrictHeaps}. Active in \cite{PadovanHeaps} is connected with parent pointer reduction what we are not concerned now. Active in \cite{FibonacciHeapsRevisited} is connected with rank/nonrank links which are called fair/na\"\i ve there.
There is another incompatibility among cited papers. In Padovan heaps the join of two roots of the same rank adds rightmost child making rank edge, while comparison of roots of different rank is reflected by adding leftmost child by a nonrank edge. The children order in \cite{StrictHeaps} is reversed.
The order in Padovan heaps differs from the choice from \cite{FibonacciHeaps}, let us make it compatible. So joining two roots of the same rank would result in adding leftmost child making rank edge while the nonrank edges would be added rightmost. Actually the children node order of solid children is not important and in the case of interface including \Meld, it is better to keep right side for deffered nodes, so even nonrank links of solid nodes would add leftmost child.

We would call solid node without parent connected by a rank edge a rank root.
It would be compatible with solid(active) roots of a variant mentioned at the end of \cite{StrictHeaps}, but it differs from the main variant where active roots are defined by not having active parent and node with minimum key is kept passive, while it would be solid in our structure. 

We would choose another path to reach the worst case bounds, but the high level describtion of the ideas would remain same.
We would start with our amortised analysis of heaps based on superexpensive comparison principles and we would change the implementation in such a way
the potencial changes would be proportional to the method times. One point of view is the potential is needless, but my point of view is natural potential still navigates us to better overall efficiency.

Let me rely on the amortized analysis of heaps based on superexpensive comparison principle from \cite{PadovanHeaps}. We have potential $\Phi_0$ representing number of trees (to pay for first phase of \FindMin). It must decrease to $1$ during \FindMin, and time of \FindMin\ should be $O(1)$ so we have to maintain number of trees constant. Calling \FindMin\ after each update in the data structure would suffice.

We have potential $\Phi_1$ corresponding to sum of differences of node degrees and their ranks. $\Phi_1$ is used in \DeleteMin\ to pay for increase of $\Phi_0$ exceeding $c\log n$ (corresponding to maximal possible rank). Actually having child deeper in the tree means longer the node neednot be compared to minimum. This is why we don't want to cut those children unless necessary. Instead we change the definition of $\Phi_1$ which would be good enough for its purpose as well. 
(As in BLT heaps we define bound function $b(n)=c_1\log n+c_2$, let us fix the constants $c_i$ later. Let us use the global list defining positions of nodes and let us check first 2 nodes after each \DeleteMin\ and move them to the list end. We can define $\Phi_1$ corresponding to sum of (positive) diferences of node degrees and their bounds $b(2n-p)$.) 
Even new definition of $\Phi_1$ is sufficient to pay for $\Phi_0$ increase during \DeleteMin\ exceeding $\theta(\log n)$.
Maintanance of $\Phi_1\le c\log n$ would be rather easy.

We have potential $\Phi_2$ corresponding to number of trees, but at most $c\log n$ (to pay for second phase of \FindMin). 
Maintaining $\Phi_0$ constant forces $\Phi_2$ be constant as well.

Last potential is $\Phi_3$ corresponding to number of nodes whose rank was decremented from the time it was linked to its parent. The strategy of Fibonacci heaps was rank could be decremented at most once (creating loss of node at most 1) and further decrement should result in decrementing parent rank and making the edge nonrank edge. In the worst case scenario we cannot maintain the loss at most 1 and bound the rank consolidation to constant time simultaneously.
Instead we would allow loss bigger than 1 so $\Phi_3$ becomes sum of losses. 
We would maintain $\Phi_3$ bounded by a logarithmic function, but we should prove the trees remain $(c,q)$ narrow for some $1\ge c>0$ and $2\ge q>1$.
Minimal size of rank subtree with root of rank $r$ and total loss in subtree at most $r+1$ is as well as in BLT heaps achived by tree with losses exactly at subtree root children which can be created by starting with binomial tree and $r+1$ times cutting grandchild of currently maximal rank. 
This therefore leads to the same quadratic equation and the same upper bound for $R(n)$ and $R(n)\le 6+1.2\log n$ is sufficient for our purposes.

In amortized heaps the main work is done in \FindMin\ where we temporarily use same rank identifying places. 
If we would maintain just one tree after each operation, there would be rarely incentive to create rank edges if we continue using this strategy.
Instead we would maintain roots in the same rank identifying places for later use even after they are linked to other nodes by nonrank edges. 
This goes against superexpensive comparison principle as we would later compare root keys which cannot be minimal, but the creation of rank edges have higher priority (in the worst case environment). The same root identifying places become violation list of rank roots. Actually we will see later that in the case of \Meld\ interface we would need two violation lists of rank roots. As the rank identifying places are not dealocated, each heap needs to keep it private.
This is why each heap would maintain its own list of ranks. 
We would make a compromise during \FindMin s. We would make violation lists reductions paid as much as allowed by $\Phi_0$ decrease (correspondnig to the first phase of amortised version), than we would make nonrank links as in the second phase of amortised version. Making nonrank links first and than do violation reductions would achieve same assymptotic bahaviour, but it would make more comparisons.
Internally \FindMin\ would process list of tree roots, the roots will have parent pointer implicitly \nill. \FindMin\ makes the parent pointeres of all the roots explicit.

\section{Violations and their reductions}
\vbox to 93mm{
\hbox{\kern7mm\pdfximage width 14cm {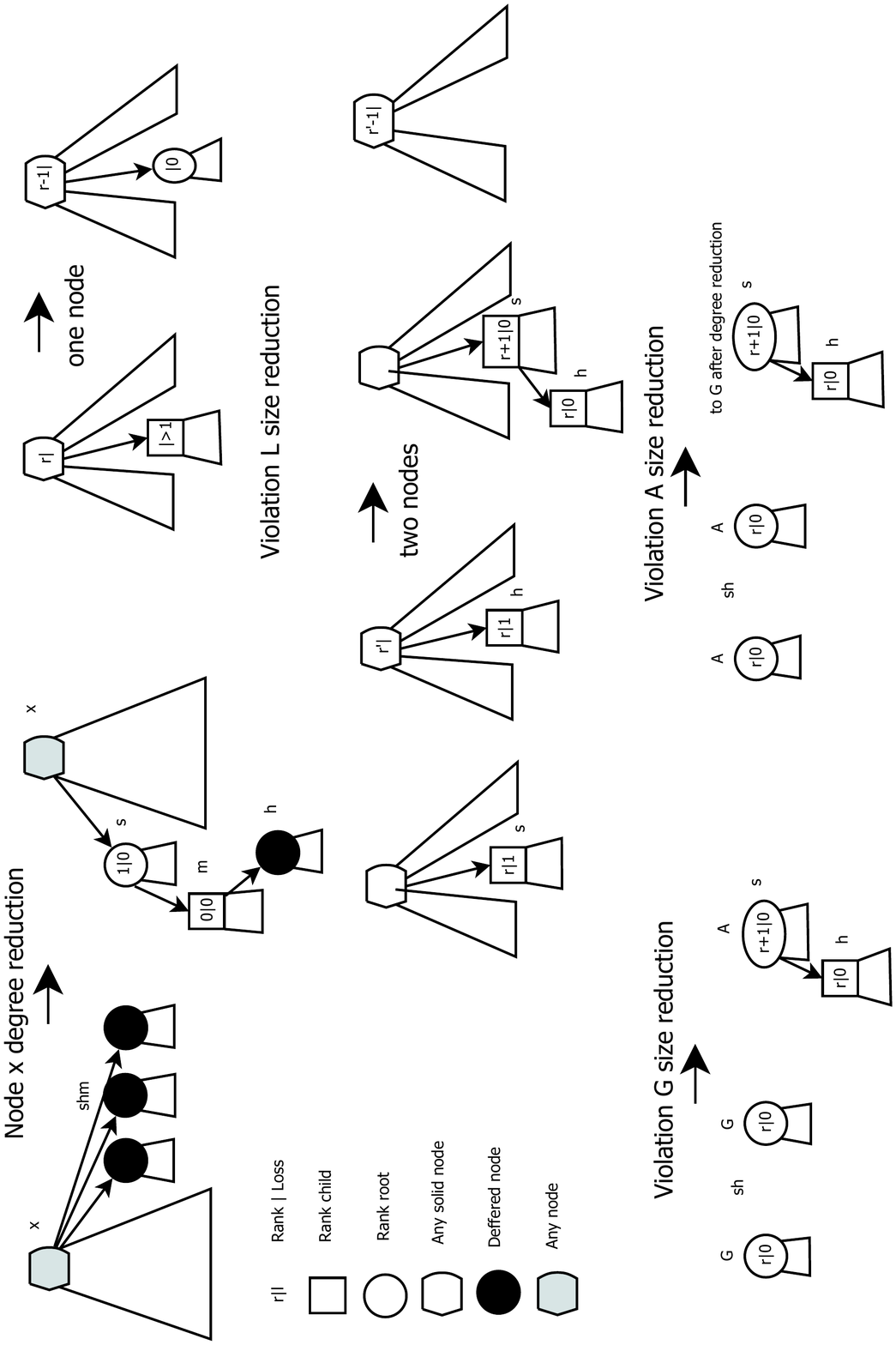}
\rlap{\smash{\pdfsave\pdfsetmatrix{0 -0.6 0.6 0}
\pdfrefximage\pdflastximage}}\pdfrestore
\hss}
\vss
\hbox{fig 1: Reductions to maintaint the heap shape}
\kern4mm
}

\begin{table}
 \begin{center}
  \caption{Effect of different transformations}
	\label{tab:eff}
	\begin{tabular}{lrrrr}
	 Changes & $|L|$ & $|A|$ & $|G|$ & Key comparisons\\
	 \hline
	 node degree reduction & $0$ &  $0$ & $+1$ & $3$  \\
	 $A$ reduction step (A)& $0$ &  $-2$ & $\le +2$ & $\le 4$ \\
	 \ - no 3 deffered chidren & $0$ & $-2$ & $+1$ & $1$ \\
	 \ - 3 deffered children reduced & $0$ & $-2$ & $+2$ & $4$ \\
	 $G$ reduction step (G)&  $0$ & $+1$ & $-2$ & $1$ \\
	 $L$ reduction step (L)& $\le -1$ & $\le 0$ & $\le +2$ & $\le 3$ \\
	 \ - one node, parent was rank root in $G$ or $A$& $-2$ & $0$ & $+1$ & $0$ \\
	 \ - one node, parent allready had loss > 0& $\le -1$ & $0$ & $+1$ & $0$ \\
	 \ - one node, parent no 3 deffered children > 0& $\le -1$ & $0$ & $+1$ & $0$ \\
	 \ - one node, parent 3 deffered children reduced & $\le -1$ & $0$ & $+2$ & $3$ \\
	 \ - two nodes, parent of $h$ is rank child& $-1$ & $0$ & $0$ & $1$ \\
	 \ - two nodes, parent of $h$ in $G$& $-2$ & $0$ & $0$ & $1$ \\
	 \ - two nodes, parent of $h$ in $A$& $-2$ & $-1$ & $+1$ & $1$ \\
	 \hline
	 $1$(A)$+1$(G)& $0$ & $-1$ & $\le 0$ & $\le 5$ \\
	 $1$(A)$+2$(G)& $0$ & $0$ & $\le -2$ & $\le 6$ \\
	\end{tabular}
 \end{center}
\end{table}

Deffered nodes would be made by the \Meld\ method.
Similarly as in BLT heaps, the nodes of smaller heap would become deffered implicitly.
Implicitly deffered nodes cannot have solid children.
Deffered nodes would be accessed during degree reductions of their parents, and during \DeleteMin s, 
when the heap root was deffered node parent or by degree reduction of the deffered node reflecting decrease of heap size when moved from the start of heap node list to its end. When the implicitly deffered node is firstly accessed, the pointer responsible for implicit deffering is removed and the node is converted to explicitly deffered. Its rank pointer is redirected to the rank 0 of current heap. All its children are deffered (either implicitly or explicitly) in this time, so there are rightmost. Explicitly deffered nodes should have rank 0, so no solid rank child is allowed, but nonrank solid children are allowed. 

As in BLT heaps, for the solid nodes with 0 loss the degree bound $b()$ is one higher than for all other nodes.
Degree reduction step on a node $x$ would be made similarly as root degree reduction on BLT heaps.
If node $x$ is implicitly deffered, it is converted to explicitly deffered.
If the rightmost 3 children are deffered, we convert them to explicitly deffered if not converted yet and we remove them from children list of $x$.
We made 3 comparisons to find order of their keys (again against the superexpensive comparisons principle) let node $s$ have the smallest, $m$ the middle, and $h$ the highest key.
We continue by making $s$ and $m$ solid.
We create rank edge making $s$ root of rank 1 having solid rank child $m$ of rank 0.
We make $h$ a deffered child of $m$, whose rank would stay 0. 
Finaly $s$ is linked as a nonrank (leftmost) child of $x$.
Degree constraints are OK for $s$ and $m$ as they become solid with loss 0.
New rank root without guaranted degree reserve was created. Degree of $x$ was reduced by 2.

Rank roots without guaranted reserve would be maintained in violation list $A$ allowing fast localisation of two nodes of the same rank if they exist in the list. Similarly active roots with guaranted reserve would be maintained in violation list $G$ allowing fast localisation of two nodes of the same rank if they exist in the list.

If $p$ was a rank root and it's rank is decremented, we know now it has guaranted reserve, so decrement of its rank could remove it from $A$ and add to $G$.

If there are two nodes of the same rank on $A$ violation list $A$ reduction step could be applied.
It links two nodes of the same rank. 
(Their keys are compared, let node $s$ be the one with smaller key while $h$ the other. We cut $h$ from its parent (nonrank edge) and put it as a rank child of $s$. This increases rank of $s$ as well as it's degree. Degree reduction is performed on $s$ what makes $s$ active root with guaranted reserve ($b()$ is enough to guareantee the degree reduction occures when there is no reserve). So both $s$ and $h$ are removed from $A$ and the active root possibly created by the degree reduction would be added to $A$. Node $s$ is added to $G$.)

If there are two nodes of the same rank on $G$ violation list $G$ reduction step could be applied.
It links two nodes of the same rank.
(Their keys are compared, let node $s$ be the one with smaller key while $h$ the other. We cut $h$ from its parent (nonrank edge) and put it as a rank child of $s$. This increases rank of $s$ as well as it's degree. Degree reduction is not performed on $s$ as there was degree reserve. So both $s$ and $h$ are removed from $G$ and $s$ is added to $A$.)

As in BLT heaps we would maintain violation list of nodes with positive loss organized in a way nodes of the same rank in the list would be easily
detected (as well as nodes with loss bigger than 1).
The loss reduction takes node $x$ with loss at least 2, it makes it nonrank child of it's parent $p$.
This creates new rank root $x$ (with loss 0 and guaranted degree reserve).
The rank of $p$ is decremented and if $p$ is a rank child, its loss is increased and new loss violation is added, but total loss was reduced by at least 1.
Degree of $p$ could have been on it's limit and the limit was decremented if the loss changed from 0 to 1, therefore degree reduction should be called on the parent if it changed loss from 0 to 1.
If $p$ was a rank root, it has to be reinserted to the same violation list it was before the rank decrement as both limit and degree did not changed.
Alternatively we take two nodes with loss 1 of the same rank and link them.
(Their keys are compared, let $h$ and $s$ be the nodes with higher and smaller keys respectively. 
 Remove $h$ from it's parent and link it under $s$ by a rank edge. 
 This reduces loss of $s$ to 0 and sets loss of $h$ to 0. 
 Original parent $p$ of $h$ decrements rank by 1, so if $p$ as a rank child new loss violation is introduced, but the total loss was reduced by at least 1. 
 Degree constraint for $s$ is OK as well as for $p$.
 If $p$ was rank root, it got degree reserve so it should be inserted to $G$ during the rank decrement.)

In table \ref{tab:eff} you can see the effect of reductions, when they take place (otherwise the bounds are not exceeded and reduction is not required).

\begin{table}
 \begin{center}
  \caption{Effect of $\ell$(L)$ + (a+\ell+\lceil g/2\rceil)$(A)$+(a+2\ell+2\lceil g/2\rceil)$(G)}
	\label{tab:kombeff}
	\begin{tabular}{lrrrr}
	 Changes & $|L|$ & $|A|$ & $|G|$ & Key comparisons\\
	 \hline
	 $\ell$(L)& $-\ell$  & $0$ & $\le +2\ell$ & $\le 3\ell$ \\
	 $(\ell+\lceil g/2\rceil)$(A)$+2(\ell+\lceil g/2\rceil)$(G)& $0$ & 0 & $\le -2\ell-g$ & $\le 6\ell+6\lceil g/2\rceil$ \\
	 $a$(A)$+a$(G)& $0$ & $-a$ & $\le 0$ & $\le 5a$ \\
	 \hline
	 total & $-\ell$ & $\le -a$ & $\le -g$ & $\le 9\ell+5a+3g+1$ \\
	\end{tabular}
 \end{center}
\end{table}

So if $|L|$, $|A|$, $|G|$ could be increased by an operation by $\ell$, $a$, $g$ respectively, 
we could make $\ell$ violation list $L$ reductions, $(a+\ell+\lceil g/2\rceil)$ violation list $A$ reductions and $a+2\ell+2\lceil g/2\rceil$ violation list $G$ reductions. See the table \ref{tab:kombeff}.
This would return each of $|L|$, $|A|$, $|G|$ to either original value or to be at most maximal rank + 1 (no two items in the list with the same rank, no nodes with loss exceedenig 1). These reductions should be done to extend as possible (if any planned is possible do it, if no planned is possible you could stop) similarly as in BLT heaps.

Degree reduction gives us equilibrium bounds for node degrees providing maximal rank is $R(n)$ and all violation lists sizes do not exceed $R(n)+1$.
The node could have at most $R(n)$ rank children, at most $R(n)+1$ nonrank children of violation list $A$, at most $R(n)+1$ nonrank children of violation list $G$ and at most 2 deffered children. If it has more children, degree reduction can be performed. 
For our analysis it would be fine to define $b(2n-p)=3R(2n-p)+6$ for solid nodes with loss 0 and $b(2n-p)=3R(2n-p)+5$ for other nodes (with $p$ being position in the global list of nodes).
(In BLT paper the bound is just $R(2n-p)+c$ as there is only one violation list of active roots and passive nodes have rank 0 and active nodes have no nonrank active children.)
With estimate $R(2n-p)\le 6+1.2\log_2(2n-p)$ we got $b(2n-p)\le 24+4\log_2(2n-p)$ so $c_2=4$ and we have to plan degree reduction by 4, what corresponds to planning two degree reduction steps for a checked node.

As in BLT heaps, linking of rank roots which are nonrank children introduces situation which cannot happen when comparing only tree roots.
In the case keyes could be equal, random choice of result would allow chosing $h$ to be predecessor of $s$ resulting in broken tree and a cycle.
To prevent this we should expect keys are all different. 
If this is not guaranted from outside, solution is to generate (different) id's for key nodes and broke ties by id's comparisons.

\section{Heap structure}

Heap information contains size of heap (node count initialised to 0), pointers to start of rank list, to list of heap tree roots, to violation lists $A$, $G$, $L$ and to the list of all heap nodes. All the lists are maintained double linked, left pointers are maintaned cyclic (left of leftmost points to rightmost). This allows access of both ends in constant time as well as adding or removing of a given node. 
Violation list node has its own left and right pointers and is interconnected by pointers with represented heap node. 
List of heap tree roots uses sibling pointers maintained in the heap nodes.
All heap nodes could have pointers internally in heap nodes as well.

A rank list node $\ell$ contain reference counter, pointer to the heap information and pointers $p_\ell^A$, $p_\ell^G$, and $p_\ell^L$ to violation lists $A$, $G$, $L$ respectively. 
Assigning node rank 0 means setting pointer to the start of the rank list.
Ranks are never set to other value, there are only incremented or decremented, what means setting pointer to the corresponding neighbour in the rank list.
With each rank change reference counters are updated.
If needed, list is extended.
We increment reference counter of the previous end of the list when it happens.
Whenever reference counter on (last) node on the list returns to 0, we shorten the list, what decrements reference counter of new list end.
When the rank 0 reference counter is 0 at the same time as heap size $<0$, the heap information is discarded.

Whether the node is rank child, nonrank child or explicitly deffered is maintained in the node state, but this is overriden by being a heap tree root or being implicitly deffered.

Each node which points through rank node to heap information with size $<0$ is implicitly deffered. 
It could be made explicitly deffered by removal from a violation list it (possibly) appears at, setting corresponding state and setting rank to 0 by pointing to current heap rank list start (the reference counter for original rank should be decremented as well). 
The removal from a violation list of implicitly deffered nodes neednot organize the violation lists by ranks, so trivial removal suffices.

Whenever node $x$ is inserted to violation list of given type $T$, violation list node $v$ pointing back to $x$ is created.
Pointer back from $x$ to $v$ could be required later.
We could use rank pointer in $x$ to save space, but in that case it should be saved in $v$ and restored when $x$ is removed from the violation list.
Placing $v$ into the list requires some care.
If pointer $p_\ell^T$ for corresponding rank list node is \nill, it is replaced by pointer to $v$ and $v$ is added to left end of violation list of type $T$. However if pointer $p_\ell^T$ is not \nill, we have at least two nodes of the rank in the violation list. 
We check existence of a neighbour of $p_\ell^T$ of the same rank.
If it does not exist, we move $p_\ell^T$ to the right end of the violation list of type $T$.
Than (in both cases), we add $v$ to the right of $p_\ell^T$. 

Whenever we remove node $x$ from the violation list, we check if $p_\ell^T$ points to $x$.
If $p_\ell^T$ points to $x$ and its right neighbour has the same rank, we let $p_\ell^T$ point to the neighbour (otherwise we set it to \nill).
Than $x$ is removed from the list and if $p_\ell^T$ is not \nill\ it's right neighbour is checked to have same rank.
If it has the same rank, we are done. Otherwise $p_\ell^T$ node should be removed and inserted (to the list left end). 
This strategy guarantees if two nodes of the same rank exists in the list, such pair can be found on its right end.
There is slight exception in violation list of type $L$, where nodes of loss at least 2 are not organized by rank.
They are simply put to the right end and eventually removed without other changes in the list.

\section{Implementation of methods}

We will describe the methods using private blocks. Their use could be slightly optimized (for example replacing pointer with \nill\ immediately before (never accessed pointer) replacing by other value could be avoided, adding node to a violation list when it will be removed from it in the same method could be avoided as well). Decomposition into blocks makes the description easier.

Before a public method is called, the violation list sizes are at most $r+1$ where $r$ is maximal possible rank.
Whenever violation list size is changed during a method, we remember the change. 
\FindMin\ adds to plan enough violation reductions and performes planned reductions to the extend possible to restore the list sizes to be each at most $r+1$, as was described in the previous two sections.
Each other public method calls \FindMin\ and does not introduce new violations after the return.

Similarly as in BLT heaps whenever we decrement size of the heap, we two times remove first node $f$ of the list of heap nodes (if it exists), we make two degree reductions on $f$ and put $f$ to the end of the list. This makes the degree constraints to hold for all nodes of the heap (assuming they have held prior to the decrement).

Whenever we decrement rank of a node $p$, violation list should be made up to date. 
We should know if decrement is done by $\alpha$) rank child removal or $\beta$) rank child conversion to nonrank child.
In the case $p$ is rank root, it must have been in either $A$ or $G$ violation list.
It should be removed from the violation list and added to position corresponding to decremented rank in $\alpha$) violation list $G$ if the degree was reduced $\beta$) to the same violation list it was before if the degree was preserved.
In the case $p$ is not rank root, it's loss is increased. 
If loss of $p$ becomes 1, it is put to violation list $L$ according it's rank. If the loss was 1, it is removed from position corresponding to the original rank and it is added among nodes with loss at least 2. Only in the case loss of $p$ was at least 2, the increment of loss is not reflected in change on a violation list.

Whenever we increment rank of a rank root $p$, it must have been in either $A$ or $G$ violation list. It should be removed from the violation list and added to position corresponding to incremented rank in the other violation list. If it was in $G$, it is added to $A$ and increment ends.
If it was in $A$, it is added to $G$ and degree reduction on $p$ is called, what finishes the rank increment.

Removal of a child $c$ of parent $p$ means following: In all cases the parent pointer of $c$ would be set to \nill\ and $c$ would be removed from the
children list of $p$ and added to the list of heap tree roots. If $c$ was a rank child, rank of $p$ is decremented\footnote{by the already described method}.

To link two solid nodes means comparing their keys, let node $s$ be the one with smaller key while $h$ the other. 
If $h$ had no parent, it is simply removed from its sibling list.
Otherwise removal of a child $h$ of its parent is invoked\prevfootnotemark.
If $h$ was rank root, it is removed from the violation list $A$ or $G$ it was in.
Node $h$ is added as a solid (therefore as leftmost) child of $s$ marking $h$ rank child if the nodes had equal rank and nonrank otherwise.
If a rank child was added, rank of $s$ should be incremented\prevfootnotemark.

\MakeHeap\ inicializes the heap structure.

\Insert($k$)\ creates new solid node $x$ with key $k$, rank 0, no parent and no child. 
It increments the heap size in the heap information without side effects.
It adds $x$ as a new root to the list of heap tree roots and invokes \FindMin.
$x$ is returned for further references. 

\FindMin\ traverses nodes of heap tree roots list and makes their parent pointers explicitly to \nill. It counts the roots meanwhile and converts implicitly deffered roots to explicitly deffered and (even new) explicitly deffered to solid, the newly solid roots are added to violation list $G$. A root which was already solid is checked to be in either $A$ or $G$ violation lists. If not, it is inserted to the violation list $A$.
Let there were $k$ roots. Add to plan $k$ violation list $A$ reductions and $k+1$ violation list $G$ reductions and do reductions to extend as possible. 
This finishes the first phase.
As $|L|$, $|A|$, $|G|$ could be increased by \FindMin\ by $0$, $a$, $g$ respectively, with $a+g\le k$, the violation sizes would remain in bounds.
Than traverse the heap tree roots leftwise linking two neighbouring roots interlaced with steps to left in the circular list (to link the roots as even as possible).
The phase ends when only one tree remains. It's root points to minimum and it will be returned.
All violations created during second phase should create new plan of violation reductions and that plan should be done to extend as possible before the return.

\DeleteMin\ decrements size\prevfootnotemark\ in the heap information. Let $\rho$ be the only tree root. It updates pointer to the list of roots to point to the leftmost child of $\rho$. It removes $\rho$ from list of heap nodes and from the violation list it is contained in. It updates reference count in it's rank. At the end it calls \FindMin\ and discards $\rho$. 

\Decrement($x$, $k$) removes $x$ from its parent $p$\prevfootnotemark\ if such parent exists.
Than in all cases it updates key at node $x$ to $k$. It invokes \FindMin\ at the end. (We could maintain keys directly in the heap nodes rather to solution in BLT heaps where separate key nodes interlinked with heap nodes are proposed).

\Meld($h_1$, $h_2$) identifies smaller heap $h_S$ and larger $h_H$ by comparing sizes in the heap informations (call with a heap with size $<0$ is invalid).
It appends list of $h_S$ nodes to start of the list of $h_H$ nodes (and sets corresponding pointer at $h_S$ to \nill). 
As position nodes of $h_S$ in the new list remain same, but the heap size at least doubles, $c_2\log_2(2n-p)$ increases by $c_2>1$ so we got reserve 1 in degree bounds so we could make solid node with loss 0 of $h_S$ deffered node of $h_H$ without violating degree constraint bounds (for other nodes of $h_S$ it is even more obvious).
It stores sum of the sizes in the heap $h_H$ informations and sets size to $-1$ in $h_S$ informations, what makes all $h_S$ nodes implicitly deffered. 
It appends roots of trees list of $h_S$ to the front of roots of trees list of $h_H$ (and sets them to \nill\ in $h_S$). 
Finally it invokes \FindMin\ and returns $h_H$ as a current heap.

\section{Simplification when \Meld\ is not needed}
As deffered nodes are created only by \Meld\ method, there will be no deffered nodes in the heap at all.
Therefore all nonrank nodes will be rank roots. 
Their number is limited by their maintenance in violation lists by $2R(n)+2$.  
This makes $\Phi_1$ bounded by $O(\log n)$ as well as each node degree.
The degree reduction is impossible and it is not needed at all, all nodes have implicitly degree reserve,
so there is no need to maintain rank roots in two different violation lists and one violation list say $A$ suffices.
As deffered nodes would not be introduced, there is no need for structure supporting implicit deffering with reference counts to allow dealocation of its parts.
The global node list to organize degree reductions is not needed as well.
So the only support needed are the two volation lists $A$ and $L$. 
If there are no deffered nodes, I would prefere inserts of nonrank nodes rather to right end of children lists for aesthetic reasons.

\begin{table}
 \begin{center}
  \caption{Effect of different transformations for no \Meld\ variant}
	\label{tab:effNoDeffered}
	\begin{tabular}{lrrr}
	 Changes & $|L|$ & $|A|$ & Key comparisons\\
	 \hline
	 $A$ reduction step (A)& $0$ &  $-1$ & $1$ \\
	 $L$ reduction step (L)& $\le -1$ & $\le +1$ & $\le 1$ \\
	 \ - one node, parent was rank root& $\le -2$ & $+1$ & $0$ \\
	 \ - one node, parent was not rank root& $\le -1$ & $+1$ & $0$ \\
	 \ - two nodes& $\le -1$ & $0$ & $1$ \\
	\end{tabular}
 \end{center}
\end{table}

The table of violation reduction steps would simplify as shown in \ref{tab:effNoDeffered}. Plan to reduce $\ell$ and $a$ violations of types $L$ and $A$ is shown in table \ref{tab:kombeffNoDeffered}.

\begin{table}
 \begin{center}
  \caption{Effect of $\ell$(L)$ + (a+\ell)$(A) for no \Meld\ variant}
	\label{tab:kombeffNoDeffered}
	\begin{tabular}{lrrrr}
	 Changes & $|L|$ & $|A|$ & Key comparisons\\
	 \hline
	 $\ell$(L)& $\le -\ell$  & $\le +\ell$ & $\le \ell$ \\
	 $(a+\ell)$(A)& $0$ & $\le -a-\ell$ & $a+\ell$ \\
	 \hline
	 total & $\le -\ell$ & $\le -a$ & $\le 2\ell+a$\\
	\end{tabular}
 \end{center}
\end{table}

\section{Concluding remarks}
There is rather big overhead in organizing violation lists by ranks. I would recomend slightly different strategy then described so far.
When new violation is added to the violation list, plan to do violation reductions steps is updated. 
If there is exactly one node of the same rank in the list, I would rather made planned violation reduction step immediately rather to reorganization of the list. 
Of course the violations created by the reduction steps should not affect the plans and when there are no more violation reductions planned we could be forced to create new pair of violating nodes of the same rank in the violation list so moving them to the right end.
I bet this caching of creating violation list nodes and of moving pairs to right end would reduce the overhead significantly.

\section{Summary}
We have shown a variant of worst case heaps not losing information by repeated linking of heap nodes under the heap roots could be implemented.
It is simpler than the main variant presented in BLT heaps according some aspects, but complicated in other aspects.
(No need for separation of keys from nodes, we bet to have better optimized number of comparisons in amortized sense.
Main drawback is we need 3 violation lists rather to 2, and because we have allowed nonrank children even to nodes with high rank, the children list size upperbound become roughly 3 times higher than those of the BLT ones, fortunately it is sufficient to do at most 2 rank reduction steps in two checked nodes to compensate for heap size decrement. The bigger upper limit for number of children is compensated in minimizing comparisons not reflected in heap edges.)

For the interface without \Meld\ two violation lists are sufficient.
The only remaining drawback is the worstcase bounds for degrees are roughly twice the bounds of the BLT heaps. 
The main advantage is the bounds are maintained implicitly and no global list with overhead of node bound checking is required at all.

For the worst case interface of \Decrement\ heaps (without \Meld) these are the fastest and simplest published heaps so far (according to my current knowledge).


\bibliography{sewc}
\end{document}